\documentclass [12pt]{article}
\usepackage {graphicx}
\usepackage[american]{babel}

\begin{document}

\title{The influence of noise on two- and three-frequency quasi-periodicity in a simple model system}

\author{A.P. Kuznetsov, S.P. Kuznetsov and Yu.V. Sedova}

\maketitle
\begin{center}
\textit{Kotel'nikov's Institute of Radio-Engineering and Electronics of RAS, \\
Saratov Branch, Zelenaya 38, Saratov, 410019, Russian Federation}
\end{center}

\maketitle
\begin{center}
\textit{Saratov State University, Astrachanskaya 83, Saratov, 410012, \\
Russian Federation}
\end{center}

\begin{abstract}
We discuss the effect of noise on a system with a quasi-periodicity of different dimensions. As the basic model of our research we use the simplest three-dimensional map with two-frequency and three-frequency quasi-periodicity. Modification of the dynamical regimes at the influence of noise is considered with the help of Lyapunov chart method. The transformation of Lyapunov exponents plots characteristic for the quasi-periodic Hopf bifurcation of 3-torus birth at the presence of noise is examined.
\end{abstract}

\section{Introduction}
The effect of noise on various dynamical regimes is an important fundamental problem, since in real physical systems the presence of noise is unavoidable \cite{1,2,3}. For systems with the possibility of chaos, noise destroys fine details of fractal structure in the phase space and parameter space, sometimes significantly modifying the observed picture. One of the interesting and less studied questions is the case of systems with quasi-periodic dynamics. At discussion of this problem an important aspect is the possibility of quasi-periodicity with a different number of incommensurable frequencies, when in the phase space invariant tori of different dimensions are observed. Recently, new results have been obtained for autonomous systems with quasi-periodicity, including problems of quasi-periodic bifurcations \cite{4,5,6,7,8,9,10}. In this report, we will consider the effect of noise on systems with two- and three-frequency quasi-periodicity and accompanying dynamical regimes.

\section{Influence of additive noise on two-frequency quasi-periodicity. Circle map}
For comparison with further results we give briefly the simplest
illustrations of the noise influence on a system with a two-frequency
quasi-periodicity. As it is known, the base model in this case is the circle
map \cite{1,2}:

\begin{equation}
\label{eq1}
x_{n + 1} = x_n + r - \frac{K}{2\pi }\sin 2\pi x_n \,\,\,(\bmod \;1).
\end{equation}
Here $x$ is the dynamical variable (phase of oscillations), $K$ and $r$ are
the amplitude and frequency parameters. The map (\ref{eq1}) describes such phenomena
as synchronization, quasi-periodic dynamics and its destruction at
transition to chaos. In Fig.1a we can see numerically calculated Lyapunov
chart in the ($r$,$K)$ parameter plane for the map (\ref{eq1}) on which the regions
of periodic regimes P, quasi-periodic two-frequency regimes $T_2 $ and chaos
C are indicated in different colors. (For the periodic regime Lyapunov
exponent is negative, for the quasi-periodic is zero and for chaotic is
positive one.) We can see the classical picture of Arnold's tongues immersed
in the region of quasi-periodicity. Above the critical line $K=1$
tongues overlapping occurs in the system, and chaotic behavior becomes possible \cite{1,2}.

\begin{figure}[!ht]
\centerline{
\includegraphics[height=11cm, keepaspectratio]{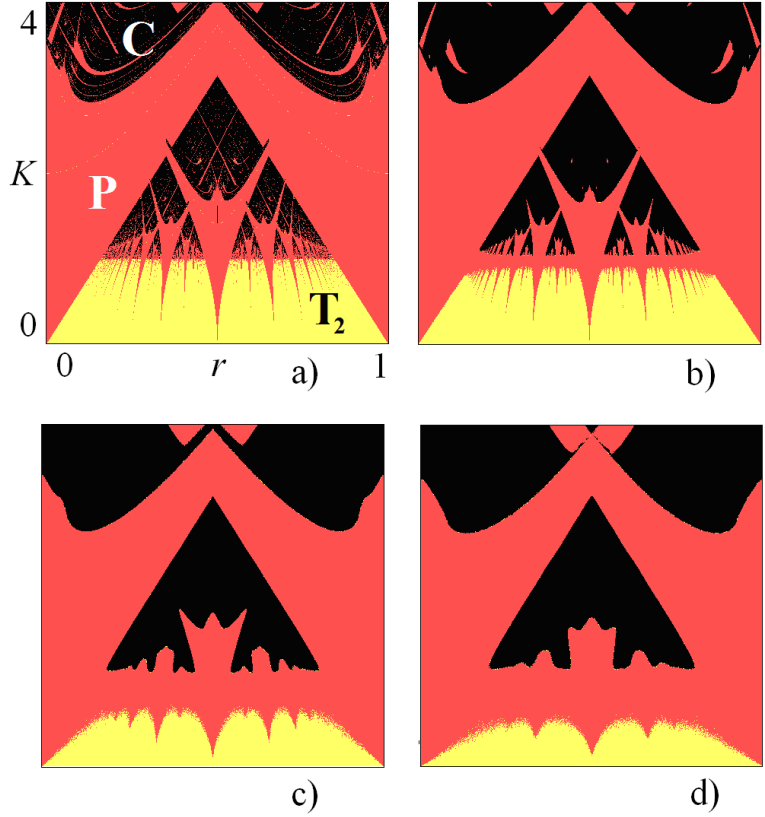}}
\caption{Lyapunov charts of the map (\ref{eq2}), a) $\gamma=0$, b) $\gamma=0.01$, c) $\gamma=0.05$, d) $\gamma=0.08$.}
\end{figure}

Let us now add a random influence to the circle map \cite{11}:
\begin{equation}
\label{eq2}
x_{n + 1} = x_n + r - \frac{K}{2\pi }\sin 2\pi x_n \, + \gamma \xi _n
\,\,\,\,\,(\bmod \;1).
\end{equation}
Here $\gamma$ is the noise intensity, and $\xi _n $ is a random
sequence of values with zero mean $\left\langle {\xi _n } \right\rangle = 0$
and constant standard deviation $\sigma = \sqrt {\left\langle {\xi _n^2 }
\right\rangle } $. For numerical calculations we use computer generated
quantities $\xi _n $ uniformly distributed over the interval $\left[ { -
0.5;0.5} \right]$. Note that if the noise amplitude is small and we examine
dynamics of the model on large time scales, then the concrete form of the
probability distribution for $\xi _n $ apparently will be not essential.

In Fig.1b,c,d Lyapunov charts corresponding to noise of different
intensities are displayed. In the presence of noise the periodic or
quasi-periodic dynamics is not realized in the exact sense, but the
structure of the typical regions on Lyapunov charts remains visible.
Therefore, we can talk about a "noisy periodic regime" when the Lyapunov
exponent is negative; about the "noisy quasi-periodic regime" when it is
close to zero; or about the "noisy chaotic regime" if Lyapunov exponent is
positive. Lyapunov charts allow us to distinguish these regimes visually.
The next conclusions follow from analysis of Lyapunov charts (Fig.1b,c,d):
\begin{itemize}
  \item At low noise intensities the structure of Arnold's tongues is
qualitatively preserved, and at large intensities it is destroyed. In the
region of small values of parameter $K$ ``relatively quasi-periodic
regimes'' dominate.

 \item In the vicinity of the critical line there is a band of ``relatively
periodic regimes'' expanding with increasing of noise intensity.

 \item Above the critical line periodic regimes are destroyed and replaced by
``relatively noisy chaotic regimes''.
\end{itemize}

Note that in the region of chaos fine details disappear, but in general this
area does not enlarge with increasing of noise amplitude, but rather reduces
in size.

\section{Influence of additive noise on three-frequency quasi-periodicity. Torus map}
Now let us consider the case of three-frequency quasi-periodicity. For its
analysis, we use the discrete model presented recently in \cite{12}. It is
generated by finite difference discretization of flow equations describing
one of the simplest generators of quasi-periodic oscillations \cite{8,9} and this
is the simplest model with the required properties. This discrete model -
the torus map - has the next form \cite{12}:
\begin{equation}
\label{eq3}
\begin{array}{l}
 x_{n + 1} = x_n + h \cdot y_{n + 1} , \\
 y_{n + 1} = y_n + h \cdot \left( {(\lambda + z_n + x_n^2 - \beta x_n^4 )y_n
- \omega _0^2 x_n } \right), \\
 z_{n + 1} = z_n + h \cdot \left( {b(\varepsilon - z_n ) - ky_n^2 } \right),
\\
 \end{array}
\end{equation}
where $x,y,z$ are dynamical variables, $\lambda ,\beta ,\omega _0
,b,\varepsilon ,k$ are the set of control parameters (for details, see
\cite{8,9,12}), $h$ is the step of discretization.

In the center of Fig.2 Lyapunov chart in the $\left( {\beta ,\lambda }
\right)$ parameter plane for the map (\ref{eq3}) is shown. Periodic regimes P are
marked by red, two-frequency quasi-periodic regimes $T_2 $ - by yellow,
three-frequency quasi-periodicity $T_3 $ - by blue, chaos $C$ - by black and
hyper-chaos \textit{HC} - by lilac color. All listed regimes were determined by the
values of Lyapunov exponents $\Lambda _i $ in accordance with their
signature:

1) P: $\Lambda _1 < 0,\,\Lambda { }_2 < 0,\;\Lambda _3 < 0$,

2) T$_{2}$: $\Lambda _1 = 0,\;\Lambda { }_2 < 0,\;\Lambda _3 < 0$,

3) T$_{3}$: $\Lambda _1 = 0,\;\Lambda { }_2 = 0,\;\Lambda _3 < 0$,

4) C: $\Lambda _1 > 0,\;\Lambda { }_2 < 0,\;\Lambda _3 < 0$,

5) HC: $\Lambda _1 > \;\Lambda { }_2 > 0,\;\Lambda _3 < 0$.

Lyapunov chart in Fig. 2 has the following features. The line of
quasi-periodic Hopf bifurcation QH separates the regions of two- and
three-frequency quasi-periodicity. From this line the bands of two-frequency
regimes originate into the domain of three-dimensional tori, these bands are
limited by lines of saddle-node bifurcations of two-frequency tori. Inside
these bands the transverse streaks of periodic regimes - exact resonances -
are built in. In the main region of two-frequency regimes periodic
resonances are possible, the widest one corresponds to 10-period cycle. Also
in Fig. 2 at certain points of parameter plane we give examples of phase
portraits and Fourier spectra illustrating the characteristic types of
regimes.

\begin{figure}[!ht]
\centerline{
\includegraphics[height=10cm, keepaspectratio]{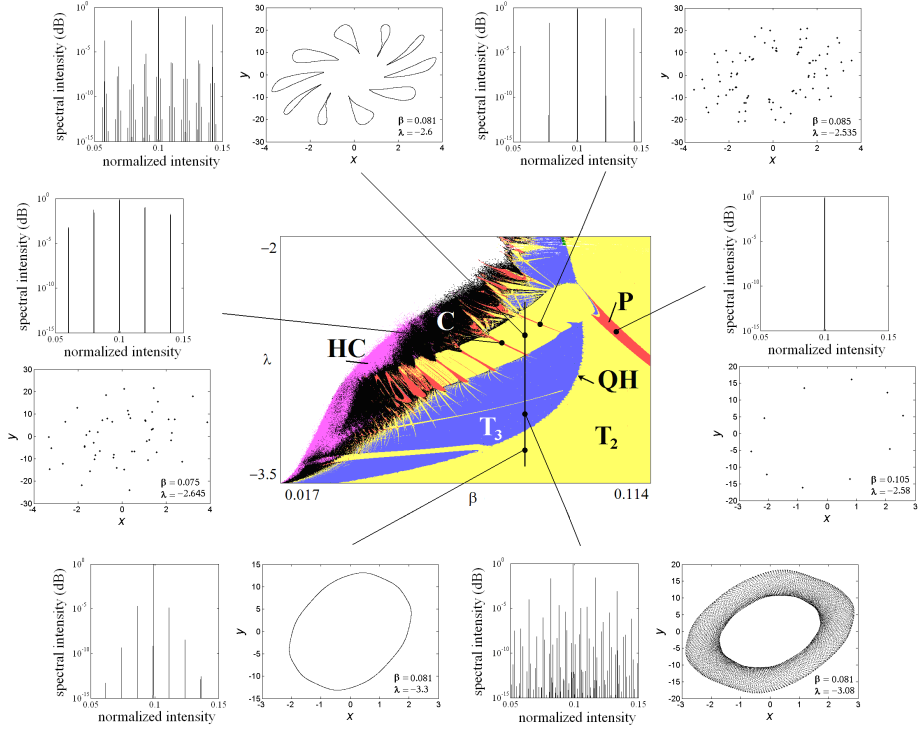}}
\caption{Lyapunov chart in the $\left( {\beta ,\lambda } \right)$ parameter
plane for the torus map (central diagram). Fourier spectra and phase
portraits are presented at several representative points. Discretization
parameter $h = 0.1$, others parameters $b = 1$, $\varepsilon = 4$, $k =
0.02$, $\,\omega _0 = 2\pi $.}
\end{figure}

To account for noise in torus map (\ref{eq3}) let us introduce random sequence $\xi
_n $. We assume that $\xi _n $ represent discrete-time white noise, i.e.,
elements of sequence at different steps of time are independent:

\begin{equation}
\label{eq4}
\begin{array}{l}
 x_{n + 1} = x_n + h \cdot y_{n + 1} , \\
 y_{n + 1} = y_n + h \cdot \left( {(\lambda + z_n + x_n^2 - \beta x_n^4 )y_n
- \omega _0^2 x_n } \right), \\
 z_{n + 1} = z_n + h \cdot \left( {b(\varepsilon - z_n ) - ky_n^2 } \right) + \gamma \xi _n , \\
 \end{array}
\end{equation}

\noindent
here $\gamma $ is the noise intensity.

Figure 3 demonstrates Lyapunov charts in the $\left( {\beta ,\lambda }
\right)$ parameter plane for the torus map at several increasing values of
noise intensity. As the amplitude of the noise increases, the following
characteristics are observed:
\begin{itemize}
  \item Small areas of periodic regimes inside the main two-frequency resonance band
are destroyed, but in their place relatively large island of ``noisy
periodic regime'' appears.

  \item The region of ``noisy periodic regime'' based on 10-period cycle is
preserved even in case of large noise values.

  \item Three-frequency regimes are preserved at small noise, but at a sufficiently
large value of noise they become two-frequency ones. Below we will
illustrate this transition by means of Fourier spectra.

  \item Chaotic and hyper-chaotic regimes survives.
\end{itemize}

\begin{figure}[!ht]
\centerline{
\includegraphics[height=8cm, keepaspectratio]{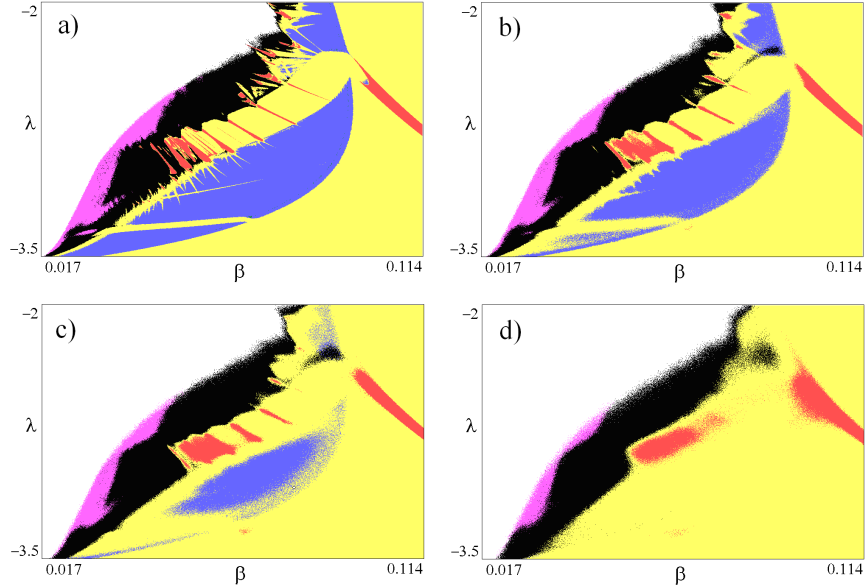}}
\caption{Lyapunov charts of model (\ref{eq4}). Noise intensity: a) $\gamma = 10^{ -
2}$, b) $\gamma = 3 \cdot 10^{ - 2}$, c) $\gamma = 5 \cdot 10^{ - 2}$, d)
$\gamma = 10^{ - 1}$.}
\end{figure}

In Fig.4 we show the hierarchy of the Fourier spectra for noisy torus map
(\ref{eq4}) with increasing noise intensity at a parameter point corresponding to
the three-frequency torus in the autonomous case. The Fourier spectrum in
the absence of noise is a discrete set of components corresponding to
incommensurable frequencies whose amplitude decreases on both sides of the
basic frequency (Fig. 4a). The peak of the maximum height is at a frequency
of 0.1, which is due to the value of the discretization parameter. With the
growth of noise, numerous satellites at the combinational frequencies
disappear (Fig. 4b-d). At noise intensity $ \approx 10^{ - 1}$ Fourier
spectrum becomes similar to the spectrum of a two-frequency quasi-periodic
oscillations (Fig. 4e). These numerical illustrations are useful because
they can be directly compared with spectra obtained as a result of physical
experiment.

\begin{figure}[!ht]
\centerline{
\includegraphics[height=8cm, keepaspectratio]{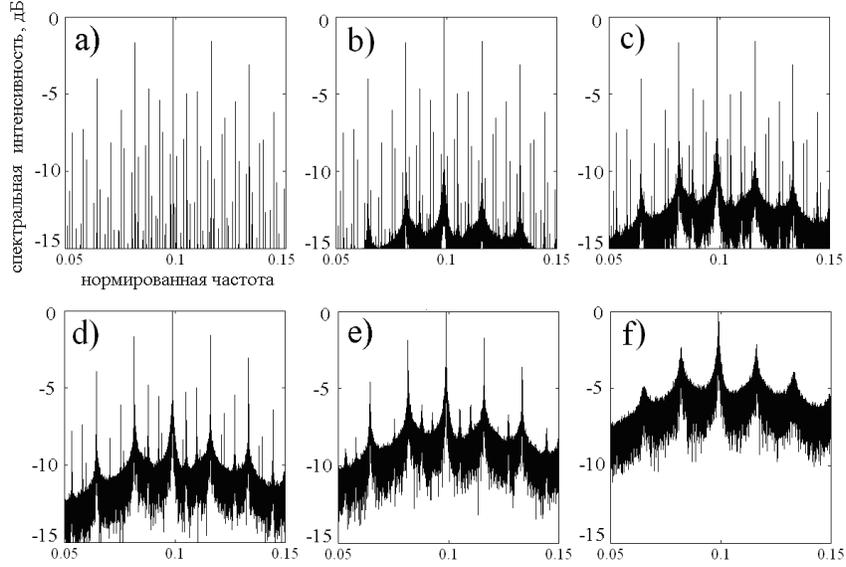}}
\caption{Fourier spectra of model (4) at point with coordinates $\beta =
0.081$, $\lambda = - 3.08$, $b = 1,\,\,\varepsilon = 4,\,\,k =
0.02,\,\,\omega _0 = 2\pi $. Noise intensity $\gamma$ is: a) 0, b) $10^{ - 5}$, c) $10^{
- 4}$, d) $10^{ - 3}$, e) $10^{ - 2}$, f) $10^{ - 1}$.}
\end{figure}

Now we discuss the influence of noise on the quasi-periodic Hopf bifurcation
\textit{QH} transforming 2-torus to 3-torus. For this purpose in Fig.5 we demonstrate
the effect of noise on the plot representing dependence of Lyapunov
exponents spectrum on parameter $\lambda $. (We move along the line $\beta =
const$ marked in Fig. 2. This line crosses the region of the three-frequency
quasi-periodicity from bottom to up). At first, let us discuss the form of
plot in the case $\gamma = 0$. At the QH point a two-frequency torus
($\Lambda { }_1 = 0)$ undergoes bifurcation. As can be seen from Fig.4a, the
attribute of this bifurcation is equality of second and third exponents
$\Lambda { }_2 = \Lambda _3 $ below bifurcation threshold. At the
bifurcation point, both these exponents vanish. Beyond the bifurcation
point, the exponents do not coincide: the second exponent is equal to zero
$\Lambda { }_2 = 0$, and the third one becomes negative $\Lambda _3 < 0$. Now
$\Lambda _1 = \,\Lambda { }_2 = 0$ and three-frequency torus is arising.
This is a quasi-periodic Hopf bifurcation QH. Its distinguishing feature is
the requirement of coincidence of two exponents below the bifurcation point
\cite{6}.

When low-intensity noise is added (Fig. 5b), the main identification feature
of this bifurcation remains: second and third exponents are equal to each
other $\Lambda { }_2 = \Lambda _3 $. However, now the point of bifurcation
is slightly blurred. This fuzziness is especially noticeable when the noise
intensity is high (Fig. 5c). At small value of the parameter, the exponents
visually coincide, but then with the growth of noise they become unequal to
each other. Thus the exponent $\Lambda { }_2$ grows, approaching zero, while
the exponent $\Lambda _3 $ reaches a maximum, which, however, is no longer
equal to zero. Then exponent $\Lambda _3 $ begins to decrease.

Fig. 5c illustrates also the behavior of a system with a saddle-node
bifurcation of invariant tori under the influence of noise. Now Lyapunov
exponent $\Lambda { }_2$ exhibits a ``dip'' into the negative values domain.
But $\Lambda { }_2$ and $\Lambda _3 $ are substantially not equal to each
other, which is an attribute of saddle-node bifurcation of tori in the
absence of noise \cite{6}. Note that as the noise intensity increases, small
features and the irregularity of the plots disappear (Fig. 5b, c), that
corresponds to the destruction of small resonances.

\begin{figure}[!ht]
\centerline{
\includegraphics[height=14cm, keepaspectratio]{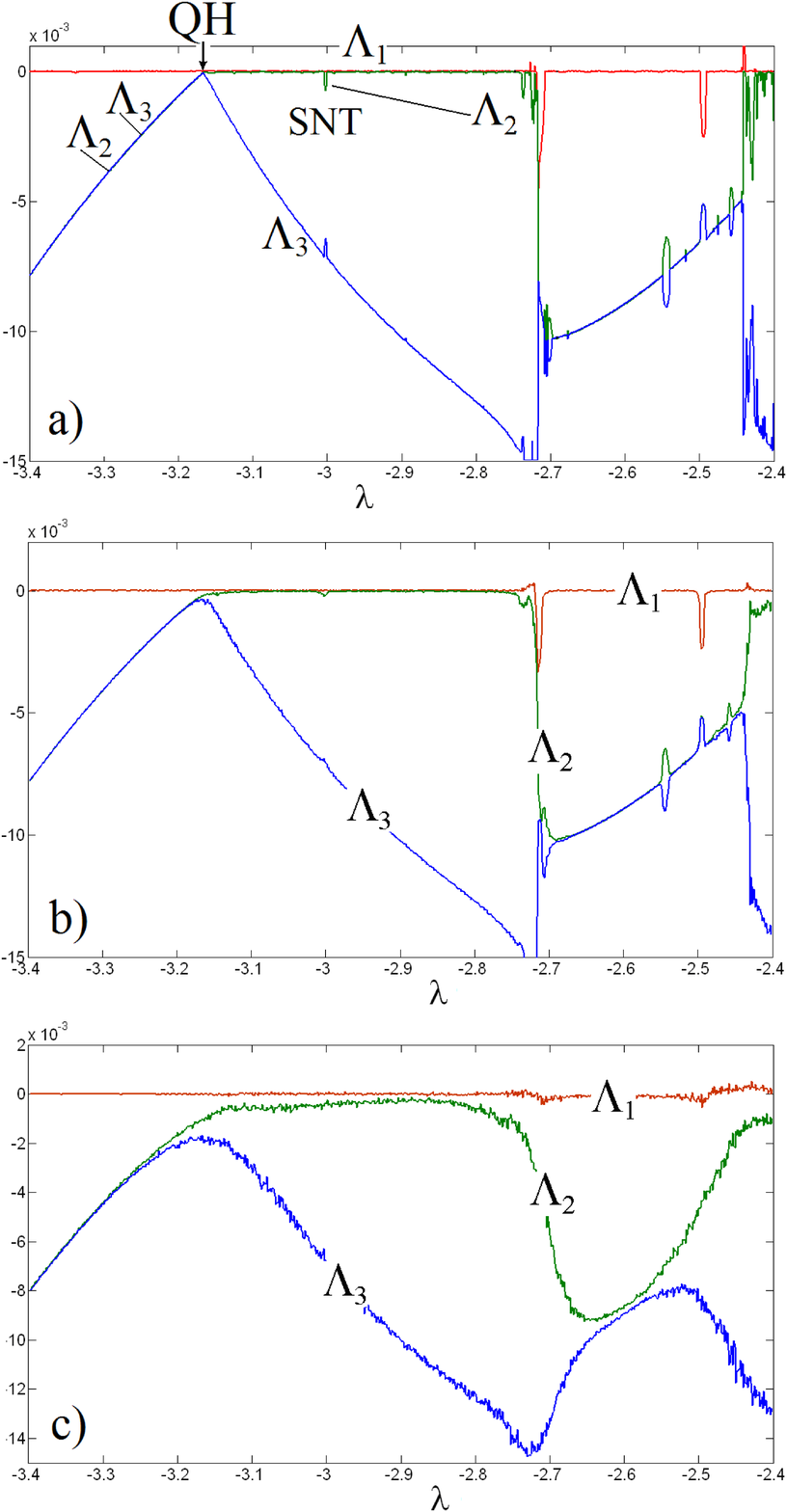}}
\caption{Lyapunov exponents of noisy torus map (4) versus parameter $\lambda
$. Noise intensity: a) $\gamma = 0$, b) $\gamma = 10^{ - 2}$, c) $\gamma = 5
\cdot 10^{ - 2}$. Parameters $b = 1$, $\varepsilon = 4$, $k = 0.02$
$\,\omega _0 = 2\pi $, discretization step $h = 0.1$. Parameter $\beta $ is
fixed, $\beta = 0.081$.}
\end{figure}

\section{Conclusion}

Thus, we consider the effect of noise on the simplest system with two- and
three-frequency quasi-periodicity. The three-frequency quasi-periodicity is
preserved for some noise amplitudes, but then turns into a two-frequency
one. For Fourier spectra this process evolves according to the scenario of
"smearing" of corresponding spectral components by noise components.
Quasi-periodic bifurcations under influence of noise occupy certain
intervals in the parameter, but their main classification characteristics
(equality or not) of corresponding Lyapunov exponents at a qualitative level
are preserved.

\textit{The research was supported by Russian Foundation for Basic Research RFBR (grant No 15-02-02893).}


\begin{thebibliography}{12}
\bibitem{1}
H.G. Schuster and W. Just {\it Deterministic chaos: an introduction}, (John Wiley {\&} Sons, 2006).

\bibitem{2}
A. Pikovsky, M. Rosenblum and J. Kurths {\it Synchronization: A Universal Concept in Nonlinear Sciences}, (Cambridge University Press, 2001).

\bibitem{3}
V.S. Anishchenko, V.V. Astakhov, A.B.Neiman, T.E. Vadivasova and L. Schimansky-Geier {\it Nonlinear dynamics of chaotic and stochastic systems: tutorial and modern developments}, (Springer Science \& Business Media, 2007).

\bibitem{4}
V.S. Anishchenko, S.M. Nikolaev and J. Kurths {\it Phys Rev E} {\bf 76} (2007) 046216.

\bibitem{5}
R. Vitolo, H. Broer and C. Sim\'{o} {\it Nonlinearity} {\bf 23} (2010) 1919-1947.

\bibitem{6}
H. Broer, C. Sim\'{o} and R. Vitolo {\it Regular and Chaotic Dynamics} {\bf 16} (2011) 154-184.

\bibitem{7}
V. Anishchenko, S. Nikolaev and J. Kurths {\it Phys Rev E} {\bf 73} (2006) 056202.

\bibitem{8}
A.P. Kuznetsov, S.P. Kuznetsov, E. Mosekilde and N.V. Stankevich {\it The European Physical Journal Special Topics} {\bf 10} (2013) 2391-2398.

\bibitem{9}
A.P. Kuznetsov, S.P. Kuznetsov and N.V. Stankevich {\it Communications in Nonlinear Science and Numerical Simulation} {\bf 15} (2010) 1676-1681.

\bibitem{10}
A.P. Kuznetsov, S.P. Kuznetsov, I.R. Sataev and L.V. Turukina {\it Phys Lett A} {\bf 377} (2013) 3291-3295.

\bibitem{11}
A.P. Kuznetsov, S.P. Kuznetsov and J.V. Sedova {\it Physica A} {\bf 359} (2006) 48-64.

\bibitem{12}
A.P. Kuznetsov and Yu.V. Sedova {\it International Journal of Bifurcation and
Chaos} {\bf 26} (2016) 1630019 (12 pages).

\end{thebibliography}
\end{document}